\documentstyle[preprint,aps,prd]{revtex}

\begin{document}

\draft
\preprint{}

\author{R.J.M. Covolan, A.V. Kisselev\thanks{On leave from the
Institute for High Energy Physics, 142284 Protvino, Russia} \ and M.S. 
Soares  \\
\small Instituto de Fisica Gleb Wataghin, UNICAMP \\ 
\small 13083-970, Campinas, SP, Brasil}

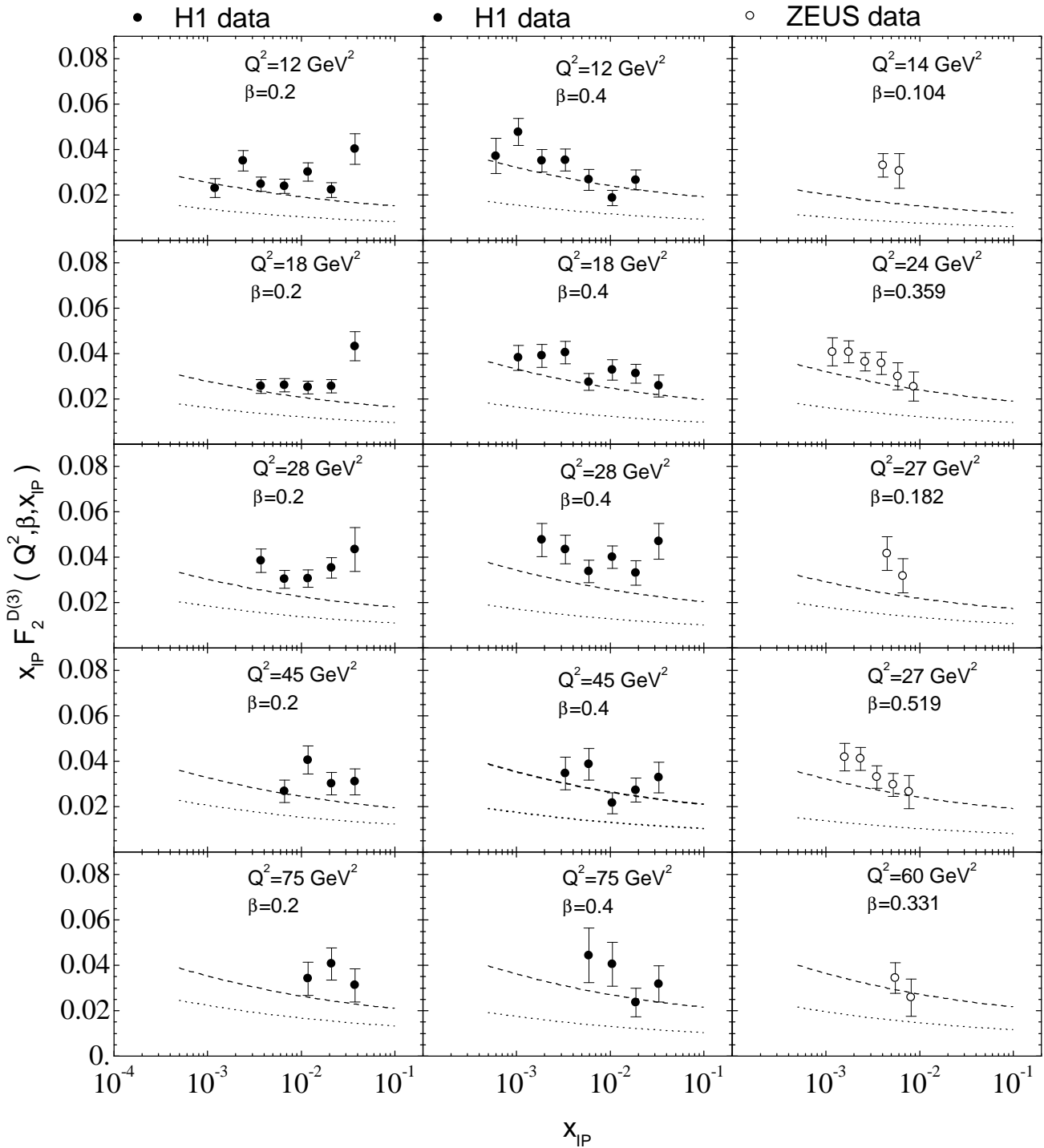
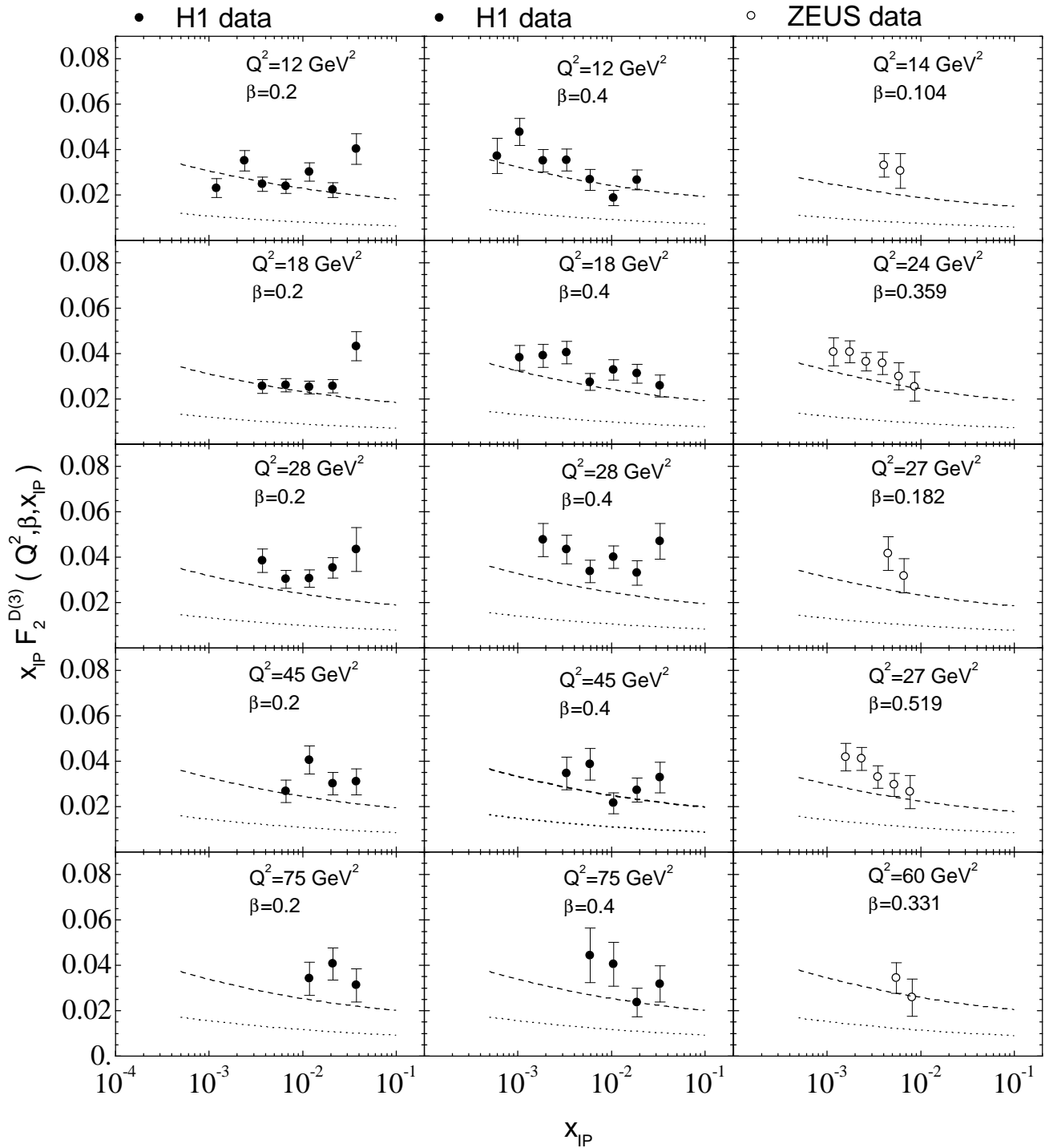
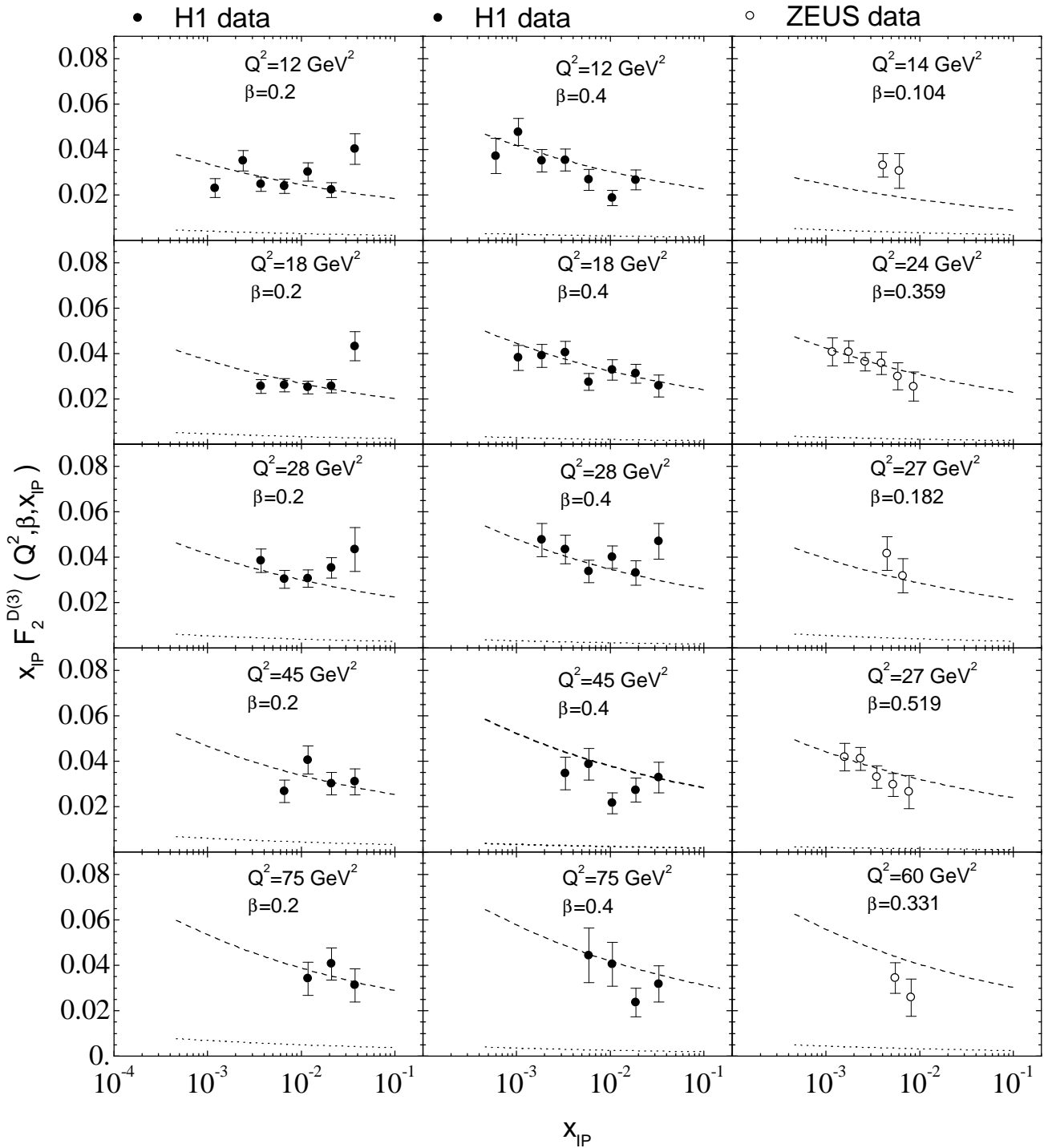
\title{Charm Contribution to the Structure Function in Diffractive Deep 
Inelastic Scattering}

\date{}

\maketitle

\begin{abstract}
\baselineskip=0.4in

The charm contribution to structure functions of diffractive deep inelastic 
scattering is considered here within the context of the Ingelman-Schlein 
model. Numerical estimations of this contribution are made from 
parametrizations of the HERA data. The influence of the Pomeron flux factor 
is analized as well as the effect of the shape of the initial parton 
distribution employed in the calculations. The obtained results indicate 
that the charm contribution to diffractive deep inelastic processes might 
be large enough to be measured in the HERA experiments.

\end{abstract}


\newpage

\section{Introduction}

The HERA data of deep inelastic scattering (DIS) measured in the last few 
years contain a sizeable fraction of events with a large rapidity gap in the 
forward region~\cite{H1new,ZEUSnew}. This phenomenon is present even at high 
$Q^2$ and results from a colour--singlet exchange between the dissociated 
virtual photon and the recoiling proton (or proton remnant), characterizing 
what is usually called {\it diffractive} DIS (DDIS). The measurement of DDIS 
at HERA provides an unique opportunity to study diffraction in regions in 
which perturbative QCD is applicable.

Open heavy quark production at HERA is also a subject of major interest in 
QCD phenomenology. Both the H1 and ZEUS collaborations have found the charm 
component of the structure function, $F_2^{(c)}(x,Q^2)$, to be a large 
fraction of $F_2(x,Q^2)$ at small $x$~\cite{H1charm,ZEUScharm}. Recently 
the first measurements of the $b \bar b$--cross section have been 
reported~\cite{HERAbottom}. Due to the higher mass of the $b$--quark, it 
is two order of magnitude smaller than the $c \bar c$--cross 
section~\cite{HERAbottom}.

For the moment most of the experimental data is for the neutral current 
structure function $F_2(x,Q^2)$. A number of theoretical estimates of 
$F_2^{(c)}(x,Q^2)$ has recently been obtained~\cite{Kisselev,Harris} 
(see also a review in Ref.~\cite{HERAbottom} and references therein). 
In the present paper we consider DDIS with open charm production and 
calculate a charm contribution to the diffractive structure function 
$F_2^D(\beta,Q^2,x_{\rm I\! P})$.  

There are two different approaches to a treatment of the charm component 
in structure functions. In one approach~\cite{Aivazis,Thorne,Martin} 
the charm is an active flavor which undergoes massless renormalization 
group (RG) evolution. We will follow another approach in which only light 
($u$, $d$, $s$) quarks and gluons are active partons and no initial state 
heavy quark lines show up in any diagram~\cite{Gluck}. It involves the 
calculation of the photon--gluon fusion process and thus is quite sensitive 
to the gluon distribution.

For the time being, the gluon distributions inside the Pomeron predicted 
by a number of models are dramatically different and they have different 
shapes. The diffractive production of open charm at HERA will, therefore, 
provide us the possibility for a direct test of the models.

The present analysis is partially based on a previous study on the Pomeron 
structure function \cite{Covolan} in which charm contribution was not 
considered. This study was mostly concerned with effects of the Pomeron 
flux factor on the evaluation of the diffractive structure fuction. Such 
effects are a central issue also in the present analysis. 

This paper is organized as follows. In Section II, we describe how to take 
into account the charm content of the Pomeron. The charm contribution to the 
diffractive structure function is evaluated in Section III, where the 
comparison  with other models is also given. Our main conclusions are 
summarized in Section IV.

\section{The Charm Content of the Pomeron}

After integration over the entire $t$ range, the  DDIS inclusive cross 
section can be written as  
\begin{equation}
\frac{d^3\sigma^D}{d\beta\ dQ^2\ dx_{\rm  I\! P}} = \frac{2\pi 
\alpha^2}{\beta\ Q^4}\ [1 + (1 - y)^2]\ F_2^{D(3)}
(\beta,Q^2,x_{\rm  I\! P}), 
\label{2}
\end{equation}
where the contribution due to longitudinal structure function, 
$F_L^{D(3)}$, has been neglected since it is expected to be small. 
Here the following kinematic variables are used to describe DDIS 
(in addition to usual DIS variables $x$, $Q^2$, $y$, and $W$): 
\begin{equation}
x_{{\rm I\! P}} \simeq \frac{M_X^2 + Q^2}{W^2 + Q^2} 
\label{4a}
\end{equation} 
and
\begin{equation}
\beta \simeq \frac{Q^2}{M_X^2 + Q^2}, 
\label{4b}
\end{equation}
where $M_X$ is the invariant mass of the diffractive system. The kinematical 
variable $x_{\rm  I\! P}$ defined in Eq.~(\ref{4a}) can be interpreted as 
the fraction of the proton momentum transferred to the Pomeron, while 
$\beta$, given by Eq.~(\ref{4b}), may be considered as the momentum fraction 
of the Pomeron carried by the quark coupling to the photon. To simplify the 
notation, in what follows we will often write $F_2^D$ instead of 
$F_2^{D(3)}(\beta,Q^2,x_{\rm  I\! P})$.

In a fit to the full data sample, H1 Collaboration has found that
a description of  $F_2^{D}$ that considers only diffractive exchange 
requires a $\beta$--dependent Pomeron intercept. However, this 
factorization breaking may be explained by introducing secondary  
trajectories~\cite{H1new}.
So, we suggest that, in the region where Pomeron exchange is the dominant 
process,  the diffractive structure function could be expressed in a 
factorized form, 
\begin{equation}
F_2^D(\beta,Q^2,x_{\rm  I\! P}) = f_{\rm I\! P/p}(x_{\rm  I\! P})\ F_{\rm  
I\! P}^D(\beta,Q^2), 
\label{6}
\end{equation}
where  $f_{\rm I\! P/p}(x_{\rm  I\! P})$ is the integrated Pomeron flux 
factor, and $F_{\rm  I\! P}^D(\beta,Q^2)$ is the Pomeron structure 
function~\cite{Covolan,Ingelman}.

The contribution of $b$--quarks to $F_2^D$ is expected to be negligible due
to the large mass of the bottom quark (as it takes place for DIS). Thus, 
we can omit this contribution and write 
\begin{equation}
F_{\rm  I\! P}^D(\beta,Q^2) = \beta \sum_a e_a^2 
\tilde F_{\rm  I\! P}^{(a)} (\beta,Q^2), 
\label{8}
\end{equation}
$e_a$ being the electric charge of the quark $a$ ($a=u,d,s,c$). 

For $Q^2 \gg m_c^2$, where $m_c$ is the charm-quark mass, we can 
regard $u$, $d$ and $s$ quarks to be massless and put (both the quark and 
antiquark are included in the distribution  $q_{\rm  I\! P} (\beta,Q^2)$)
\begin{equation}
\tilde F_{\rm  I\! P}^{(u)}(\beta,Q^2) = \tilde F_{\rm  I\! P}^{(d)} 
(\beta,Q^2) = \tilde F_{\rm  I\! P}^{(s)} (\beta,Q^2) = q_{\rm  I\! P} 
(\beta,Q^2), 
\label{10}
\end{equation}
that results in
\begin{equation}
F_{\rm  I\! P}^D(\beta,Q^2) = \frac{2}{3}\ \beta\ q_{\rm  I\! P} 
(\beta,Q^2) + \frac{4}{9}\ \beta\ \tilde F_{\rm  I\! P}^{(c)} 
(\beta,Q^2,m_c^2). 
\label{12}
\end{equation}

Recently, a factorization theorem has been proved for 
diffractive lepton scattering off nucleons~\cite{Collins} from which 
structure functions of DDIS coincide with DIS structure functions.
Therefore, quark and gluon distributions inside the Pomeron, 
$q_{\rm  I\! P} (\beta,Q^2)$ and $g_{\rm  I\! P} (\beta,Q^2)$, obey the same 
set of RG evolution equations as quark and gluon distributions inside the 
proton do. As the observed values of $\beta$ are not too small, DGLAP 
equations~\cite{DGLAP} can be used to perform such an evolution.

In the present analysis, we suppose that charm quarks are mainly produced 
by virtual photon--gluon fusion and do not take part in the evolution of 
the light quarks. In such a case, by analogy with charm contribution to 
$F_2$~\cite{Kisselev}, we get the following equation for the 
charm contribution to DDIS structure function $F_2^D$, 
\begin{equation}
\tilde F_{\rm  I\! P}^{(c)}(\beta,Q^2,m_c^2) = \int_{Q_0^2}^{Q^2} 
\frac{dk^2}{k^2} \int_{\beta}^1 \frac{dz}{z}\ C_g(z, Q^2,k^2,m_c^2) \  
\frac{\partial}{\partial \ln k^2} g_{\rm  I\! P} 
\left( \frac{\beta}{z},k^2 \right),  
\label{14}
\end{equation}
in which $Q_0 = 2$ GeV is assumed.

Now, we isolate a similar term in $q_{\rm  I\! P} (\beta,Q^2)$ and call 
the rest of $q_{\rm  I\! P} (\beta,Q^2)$ ``direct contribution", that is 
\begin{equation}
q_{\rm  I\! P}(\beta,Q^2) = q_{\rm  I\! P}^{dir}(\beta,Q^2) + 
\int_{Q_0^2}^{Q^2} 
\frac{dk^2}{k^2} \int_{\beta}^1 \frac{dz}{z}\ C_g(z,Q^2,k^2,0)\  
\frac{\partial}{\partial \ln k^2} g_{\rm  I\! P} \left( \frac{\beta}{z},k^2 
\right). 
\label{16}
\end{equation}

Let us define the quantity
\begin{eqnarray}
\Delta \tilde F_{\rm  I\! P}^{(c)}(\beta,Q^2,m_c^2) &=& 
\int_{Q_0^2}^{Q^2} 
\frac{dk^2}{k^2} \int_{\beta}^1 \frac{dz}{z} [C_g(z,Q^2,k^2,0) 
- C_g(z,Q^2,k^2,m_c^2)] \nonumber \\
&\times& \frac{\partial}{\partial \ln k^2} 
g_{\rm  I\! P} \left( \frac{\beta}{z},k^2 \right). 
\label{18}
\end{eqnarray}

By using these definitions, from Eqs.~(\ref{12})-(\ref{18}) we obtain
\begin{eqnarray}
 F_{\rm  I\! P}^D(\beta,Q^2) &&= \frac{2}{3}\ \beta\ \left[
q_{\rm  I\! P}^{dir}(\beta,Q^2) + \int_{Q_0^2}^{Q^2} 
\frac{dk^2}{k^2} \int_{\beta}^1 \frac{dz}{z}\ C_g(z,Q^2,k^2,0)\ 
\frac{\partial}{\partial 
\ln k^2} g_{\rm  I\! P} \left( \frac{\beta}{z},k^2 \right) \right] + 
\nonumber \\ 
&& + \frac{4}{9}\ \beta\ \int_{Q_0^2}^{Q^2} \frac{dk^2}{k^2} 
\int_{\beta}^1 \frac{dz}{z}\ C_g(z,Q^2,k^2,m_c^2)\ 
\frac{\partial}{\partial 
\ln k^2} g_{\rm  I\! P} \left( \frac{\beta}{z},k^2 \right) 
\nonumber \\ \nonumber \\
&& = \frac{2}{3}\ \beta\ q_{\rm  I\! P}^{dir}(\beta,Q^2)
+ \frac{10}{9}\ \beta\ \int_{Q_0^2}^{Q^2} \frac{dk^2}{k^2} 
\int_{\beta}^1 \frac{dz}{z}\ C_g(z,Q^2,k^2,m_c^2)\ 
\frac{\partial}{\partial 
\ln k^2} g_{\rm  I\! P} \left( \frac{\beta}{z},k^2 \right) +  
\nonumber \\&& + \frac{2}{3}\ \beta\ \Delta 
\tilde F_{\rm  I\! P}^{(c)}
(\beta,Q^2,m_c^2) \nonumber \\ \nonumber \\
&& = \frac{2}{3}\ \beta\ q_{\rm  I\! P}^{dir}(\beta,Q^2)
+ \frac{5}{2}\ F_{\rm  I\! P}^{(c)}(\beta,Q^2,m_c^2)
+ \frac{2}{3}\ \beta\ \Delta \tilde F_{\rm  I\! P}^{(c)}
(\beta,Q^2,m_c^2).
\label{20}
\end{eqnarray}

It follows from Eq.~(\ref{20}) that 
\begin{equation}
F_{\rm  I\! P}^{(c)}(\beta,Q^2,m_c^2)  = \frac{2}{5} \left[ 
F_{\rm  I\! P}^D(\beta,Q^2) -  \frac{2}{3}\ \beta\ q_{\rm  I\! P}^{dir} 
(\beta,Q^2) -  \frac{2}{3}\ \beta\ \Delta \tilde F_{\rm  I\! P}^{(c)} 
(\beta,Q^2,m_c^2) \right], \label{21}
\end{equation}

An analogous difference between DIS structure functions with and 
without charm contribution, $\Delta  F_2(x,Q^2,m_c^2)$, was calculated 
in Ref.~\cite{Kisselev}, 
where it was shown that it scales at high $Q^2$. The generalization 
for DDIS is straightforward and the result  
(up to corrections $\mbox{\rm O}(m_c^2/Q^2))$ reads 
\begin{eqnarray}
\Delta \tilde F_{\rm  I\! P}^{(c)} &=&  \Delta \tilde 
F_{\rm  I\! P}^{(c)}(\beta,m_c^2) \nonumber \\
&=& \int_{Q_0^2}^{\infty} \frac{dk^2}{k^2} \int_{\beta}^1 
\frac{dz}{z} \Delta C \left(\frac{m_c^2}{k^2}, z \right) 
\frac{\partial}{\partial \ln k^2} g_{\rm  I\! P} \left( \frac{\beta}{z}, 
k^2 \right). 
\label{22}
\end{eqnarray}
In $\alpha_s$ order the expression for $\Delta C$ is of the 
form\footnote{There are two misprints in formula (40) of 
Ref.~\cite{Kisselev} that were corrected in Eq.~(\ref{24}):  
the expression $P_{qg}(y)=1/2[(1-y)^2 + y^2]$ should be put within 
the curly brackets and the factor $\alpha_s/4\pi$ should be 
replaced by $\alpha_s/2\pi$.}~\cite{Kisselev}
\begin{equation}
\Delta C(v, u) =  \frac{\alpha_s}{\pi} \left\{P_{qg}(u) \ln 
\left[1 + \frac{v}{u(1 - u)} \right] - \frac{1}{2} (1 - 2u)^2 
\frac{v}{v + u(1 - u)} \right\}. \label{24}
\end{equation} 

Formula (\ref{22}) does not contradict the factorization theorem 
for DDIS~\cite{Collins}. Namely, if we put $k^2 = 0$ in 
$C_g(\beta,Q^2,k^2,m_c^2)$ as it is usually done, the main 
contribution in Eq.~(\ref{14}) is due to the region $k^2 \sim Q^2$, 
and one gets 
\begin{equation}
\tilde F_{\rm  I\! P}^{(c)}(\beta,Q^2,m_c^2) \simeq 
\int_{\beta}^1 \frac{dz}{z} \, C_g \left(\frac{m_c^2}{Q^2}, z \right) 
\, g_{\rm  I\! P} \left( \frac{\beta}{z},Q^2 \right). \label{26}
\end{equation}

On the other hand, in the difference of the diffractive structure functions, 
$\Delta \tilde F_{\rm  I\! P}^{(c)}(\beta,m_c^2)$, Eq.~(\ref{22}), the 
leading contributions cancell out. The quantity $\Delta C$ has the 
asymptotic behavior
\begin{equation}
\left. \Delta C \left(\frac{m_c^2}{k^2}, \beta \right) \right|_{|k^2| 
\rightarrow \infty} \sim \frac{m_c^2}{k^2}, 
\label{28}
\end{equation}
and the  main contribution to the integral in $k^2$ in Eq.~(\ref{22})
comes from the region $k^2 \sim m_c^2$~\cite{Kisselev}.

As can be seen from Eq.~(\ref{21}), $F_{\rm  I\! P}^{(c)}$ is 
defined via $\Delta \tilde F_{\rm  I\! P}^{(c)}$. In its turn, $\Delta 
\tilde F_{\rm  I\! P}^{(c)}$ is given by formula (\ref{22}) which contains 
a derivative of $g_{\rm  I\! P}$ in $\ln k^2$. This is related to the fact 
that we started from an exact expression for the coefficient function 
$C_g$~\cite{Kisselev} depending on both $m_c^2$ and $k^2$. Thus, one can 
expect that the charm contribution to the DDIS structure function should 
significantly be dependent on both the form and evolution of the gluon 
distribution inside the Pomeron.

For numerical estimates we shall use the quark and gluon distribution
functions which have been obtained in Ref.~\cite{Covolan} by fitting 
the data on 
$F_2^{D(3)}$ from H1 and ZEUS collaborations~\cite{H1old,ZEUSold}. 
Since  in this analysis it was assumed $N_f=3$   ($N_f$ being 
the number of flavors), we have to  rewrite Eq.~(\ref{21}) in terms of 
corresponding parton distributions, $q_{\rm  I\! P}^{(3)}$ and 
$g_{\rm  I\! P}^{(3)}$. Let us also define $q_{\rm  I\! P}^{(4)}$ 
($g_{\rm  I\! P}^{(4)}$) to be a quark (gluon) distribution for the case 
$N_f=4$.

It is useful to introduce the quantities
\begin{equation}
\Delta q_{\rm  I\! P} = q_{\rm  I\! P}^{(4)} - q_{\rm  I\! P}^{(3)} 
\label{30}
\end{equation}
and
\begin{equation}
\Delta g_{\rm  I\! P} = g_{\rm  I\! P}^{(4)} - g_{\rm  I\! P}^{(3)}. 
\label{32}
\end{equation}

It should be noted that
\begin{equation}
\Delta q_{\rm  I\! P} = - \frac{3}{2} {F}_{\rm  I\! P}^{(c)}. 
\label{34}
\end{equation}

From Eq.~(\ref{21}), one obtains
\begin{eqnarray}
F_{\rm  I\! P}^{(c)}(\beta,Q^2,m_c^2) &=& \left. \frac{2}{3} 
F_{\rm  I\! P}^D(\beta,Q^2) - \frac{4}{9}\ \beta \left[ 
q_{\rm  I\! P}^{dir}(\beta,Q^2) + \Delta \tilde 
F_{\rm  I\! P}^{(c)}(\beta,m_c^2) \right] \right|_{N_f=3} +  \nonumber \\
&+& \frac{4}{9} \int_{Q_0^2}^{Q^2} \frac{dk^2}{k^2} 
\frac{\alpha_s}{\pi} \int_{\beta}^1 {dz} \, P_{cg}(z)\ \frac{\beta}{z}\  
\Delta g_{\rm  I\! P} \left( \frac{\beta}{z},k^2 \right), 
\label{36}
\end{eqnarray}  
where the subscript $|_{N_f=3}$ means that the corresponding quantities 
in the right hand side (RHS) of Eq.~(\ref{36}) should be calculated with 
the use of the distributions $q_{\rm  I\! P}^{(3)}$ and $g_{\rm  I\! P}^{(3)}$.

The next step is to estimate the quantity $\Delta g_{\rm  I\! P}$ which 
enters into Eq.~(\ref{36}). Due to our assumption (no charm in light quark 
evolution), $q_{\rm  I\! P}^{(3)}$ and $q_{\rm  I\! P}^{(4)}$ obey one and 
the same DGLAP evolution equation~\cite{DGLAP}, 
\begin{eqnarray}
q_{\rm  I\! P}^{(n)}(\beta,Q^2) &=& q_{\rm  I\! P}^{(n)}(\beta,Q_0^2) + 
3 \int_{Q_0^2}^{Q^2} \frac{dk^2}{k^2} \frac{\alpha_s}{2\pi} 
\int_{\beta}^1 \frac{dz}{z} P_{qq}(z) \, 
q_{\rm  I\! P}^{(n)} \left( \frac{\beta}{z},k^2 \right) +\nonumber \\
&+& \int_{Q_0^2}^{Q^2} \frac{dk^2}{k^2} \frac{\alpha_s}{\pi} 
\int_{\beta}^1 \frac{dz}{z} P_{qg}(z) \, 
g_{\rm  I\! P}^{(n)} \left( \frac{\beta}{z},k^2 \right), 
\label{38}
\end{eqnarray}
for $n=3,4$. The factor $3$ in front of the first integral in the RHS of 
Eq.~(\ref{38}) is related to the number of light flavors. 

As for initial quark and gluon distributions inside the Pomeron, 
we have  $g_{\rm  I\! P}^{(4)}(\beta,Q_0^2) \neq 
g_{\rm  I\! P}^{(3)}(\beta,Q_0^2)$, while
\begin{equation}
\Delta q_{\rm  I\! P}(\beta,Q_0^2) = 0, 
\label{40}
\end{equation}
that is no intrinsic charm in the Pomeron.

If we neglect the variation of $\alpha_s$ with the change of the flavor 
number from $N_f=3$ to $N_f=4$, from Eq.~(\ref{38}) we get 
\begin{eqnarray}
\Delta q_{\rm  I\! P}(\beta,Q^2) &=& 3 \int_{Q_0^2}^{Q^2} \frac{dk^2}{k^2} 
\frac{\alpha_s}{2\pi} \int_{\beta}^1 \frac{dz}{z} P_{qq}(z) \, 
\Delta q_{\rm  I\! P} \left( \frac{\beta}{z},k^2 \right) +\nonumber \\ 
&+& \int_{Q_0^2}^{Q^2} \frac{dk^2}{k^2} \frac{\alpha_s}{\pi} 
\int_{\beta}^1 \frac{dz}{z} P_{qg}(z) \, \Delta 
g_{\rm  I\! P} \left( \frac{\beta}{z},k^2 \right). 
\label{42}
\end{eqnarray}
The QCD--evolution parameter
\begin{equation}
\xi (Q^2) = \frac{1}{2\pi b} \ln \left( \frac{\ln 
\displaystyle (Q^2/\Lambda^2)}
{\ln \displaystyle (Q_0^2/\Lambda^2)} \right),
\label{44}
\end{equation}
where $12\pi b=33 - 2N_f$, rises slowly in $Q^2$ and is numerically small 
even at rather high values of $Q^2$. For instance, for $Q_0=2$ GeV and 
$\Lambda = 0.2$ GeV we find $\xi (10^2 \ \mbox{\rm GeV}^2) \simeq 0.13$, 
$\xi (10^3 \ \mbox{\rm GeV}^2 ) \simeq 0.24$. In particular, it enables 
one to solve the DGLAP equations by using an expansion in the parameter 
$\xi$~\cite{Field}.

From all said above, we obtain (up to small corrections $\mbox{\rm O} 
(\xi^2)$) 
\begin{equation}
\Delta q_{\rm  I\! P}(\beta,Q^2) \simeq \int_{Q_0^2}^{Q^2} \frac{dk^2}{k^2} 
\frac{\alpha_s}{\pi} \int_{\beta}^1 \frac{dz}{z} P_{qg}(z) \, 
\Delta g_{\rm  I\! P} \left( \frac{\beta}{z},k^2 \right). 
\label{46}
\end{equation}

Due to the fact that $C_g(z,Q^2,k^2,m_c^2)$ has no large logarithms, at
$k^2 \simeq Q^2$ (see an explicit expression for $C_g(z,Q^2,k^2,m_c^2)$ in
Ref.~\cite{Kisselev}), we obtain from Eq.~(\ref{14}) in the leading
logarithmic approximation (LLA) the expression 
\begin{equation}
F_{\rm  I\! P}^{(c)}(\beta,Q^2,m_c^2) \simeq \int_{Q_0^2}^{Q^2} 
\frac{dk^2}{k^2} \frac{\alpha_s}{\pi} \int_{\beta}^1 {dz}\ P_{cg} 
\left(z, \frac{m_c^2}{Q^2} \right) \frac{\beta}{z}\ 
g_{\rm  I\! P}^{(4)} \left( \frac{\beta}{z},k^2 \right),
\label{48}
\end{equation}
where $P_{cg}(z,m_c^2/Q^2) = -\partial C_g(z,Q^2,k^2,m_c^2)/\partial 
\ln k^2|_{k^2=Q^2}$ 
is the modified form of the $P_{qg}$ splitting function for the charm 
quark~\cite{Roberts}. 

Now let us define 
\begin{eqnarray}
\int_{\beta}^1 \frac{dz}{z}\ P_{qg}(z) \ \Delta g_{\rm  I\! P}  
\left( \frac{\beta}{z},Q^2 \right) &=& r_q \int_{\beta}^1 \frac{dz}{z}\  
P_{qg}(z) \ g_{\rm  I\! P}^{(3)} \left( \frac{\beta}{z},Q^2 \right) 
\label{50a}
\end{eqnarray}
and
\begin{eqnarray}
\int_{\beta}^1 \frac{dz}{z}\ P_{cg}(z) \ \Delta g_{\rm  I\! P}   
\left( \frac{\beta}{z},Q^2 \right) &=& r_c \int_{\beta}^1 \frac{dz}{z}\  
P_{cg}(z) \ g_{\rm  I\! P}^{(3)} \left( \frac{\beta}{z},Q^2 \right),
\label{50b}  
\end{eqnarray}
and put $r_c \simeq r_q = r$. This means that we ignore additional 
subleading $m_c^2$--dependent terms in $P_{cg}$ with respect to $P_{qg}$.
The quantity $r$ may in principle depend on both $\beta$ and $Q^2$. 
Numerical estimates have shown, however, that it is weakly dependent 
on variable $Q^2$. 

From Eqs.~(\ref{34}), (\ref{46})--(\ref{50b}), we calculate $r$ and rewrite 
$F_{\rm  I\! P}^{(c)}$ in the following form 
\begin{equation}
F_{\rm  I\! P}^{(c)}(\beta,Q^2,m_c^2) = \frac{2}{3} \frac{A\ B}{A + B},
\label{52}
\end{equation}
where 
\begin{equation}
A = F_{\rm  I\! P}^D(\beta,Q^2) - \frac{2}{3}\ \beta \left. \left[ 
q_{\rm  I\! P}^{dir} (\beta,Q^2) +
\Delta \tilde F_{\rm  I\! P}^{(c)}(\beta,m_c^2) \right] 
\right|_{N_f=3}
\label{54}
\end{equation}
and
\begin{equation}
B = \int_{Q_0^2}^{Q^2} \frac{dk^2}{k^2} \frac{\alpha_s}{\pi} 
\int_{\beta}^1 {dz}\ P_{qg}(z)\ \frac{\beta}{z}\ g_{\rm  I\! P}^{(3)} 
\left( \frac{\beta}{z},k^2 \right).
\label{56}
\end{equation}
Here $F_{\rm  I\! P}^D$ is the Pomeron structure function, while 
$q_{\rm  I\! P}^{dir}$ and $\Delta \tilde F_{\rm  I\! P}^{(c)}$
are defined in Eqs.~(\ref{16}) and (\ref{22}).

\section{The Charm Contribution to the Diffractive Structure Function}
 
In this section, we present quantitative results obtained in the present 
analysis as well as some comparison with other models.

\subsection{Results of the present analysis}

Our concern now is the calculation of the charm contribution to 
$F_2^{D(3)}(\beta,Q^2,x_{\rm  I\! P})$ in two different approaches 
and for different shapes of 
the quark and gluon distributions inside the Pomeron. In one approach 
the {\em standard} flux factor is employed, whereas in the other  
the {\em renormalized} flux factor is used (for brevity, we will refer 
to these quantities hereafter as STD and REN flux factors, respectively). 
For the former, it was assumed the Donnachie--Landshoff 
expression~\cite{Donnachie},
\begin{equation}   
f_{STD}(x_{\rm I\! P},t) = \frac{9\beta_0^2}{4\pi^2}\ [F_1(t)]^2\ 
x_{\rm I\! P}^{1 - 2\alpha(t)}
\label{58} 
\end{equation}
while the latter is determined from the procedure prescribed 
in~\cite{Goulianos}, that is 
\begin{equation}
f_{REN}(x_{\rm I\! P},t) = \frac{f_{STD}(x_{\rm I\! P},t)}{N(x_{\rm I\! 
P_{min}})}
\label{60}
\end{equation}
where 
\begin{equation}
N(x_{\rm I\! P_{min}}) = \int_{x_{\rm I\! 
P_{min}}}^{x_{\rm I\! P_{max}}}dx_{\rm I\! P}
\int^0_{t=-\infty}f_{STD}(x_{\rm I\! P},t)\ dt.
\label{62}
\end{equation}
By introducing Eq.~(\ref{58}) into Eq.~(\ref{62}) and assuming
an exponential approximation for the form factor, $F_1^2(t) \simeq
e^{b_{0}(t)}$, one obtains
\begin{equation}
N(x_{\rm I\! P_{min}}) = K\ \frac{e^{-\gamma}}{2\alpha'}\ [E_i(\gamma - 
2\epsilon\ \ln x_{\rm I\! P_{min}}) - E_i(\gamma - 2\epsilon\ 
\ln x_{\rm I\! P_{max}})],
\label{64}
\end{equation}
where $E_i (x)$ is the exponential integral, $K = {9\beta_0^2}/
{4\pi^2}$ and $\gamma = b_0 \epsilon/\alpha'$. The minimum value of 
$x_{\rm I\! P}$ is $x_{\rm I\! P_{min}} = (m_p + m_{\pi})^2/s$ for soft 
diffractive dissociation and $x_{\rm I\! P_{min}} = Q^2/\beta s$ for 
DDIS~\cite{Goulianos}.

The distributions of the quarks and gluons inside the Pomeron, 
$q_{\rm I\! P}^{(3)}$ and $g_{\rm I\! P}^{(3)}$, were obtained from 
HERA data~\cite{H1old,ZEUSold} in Ref.~\cite{Covolan} (we refer the 
reader to this paper for details). The parametrizations for each flux 
factor are described below. No sum rules were imposed on them to perform 
the fitting.
\bigskip

\noindent{\bf Fit~1}: Parametrizations obtained in with STD flux in 
which both quark and gluon distributions have a hard shape at the 
initial scale of evolution:  
\begin{eqnarray} 
3\beta\ q_{\rm I\! P}^{(3)}(\beta,Q^2_0) &=& 2.55\ \beta\ (1- \beta), 
\nonumber \\ 
\beta\ g_{\rm I\! P}^{(3)}(\beta,Q^2_0) &=& 12.08\ \beta\ (1- \beta).  
\label{66}
\end{eqnarray}

\noindent{\bf Fit~2}: Parametrizations obtained
with the STD flux; the initial distributions correspond to a super--hard
profile imposed to gluons by a delta function while quarks were left
free to change according to the data:
\begin{eqnarray} 
3\beta\ q_{\rm I\! P}^{(3)}(\beta,Q^2_0) &=& 1.51\ \beta^{0.51}\ 
(1- \beta)^{0.84}, \nonumber \\ 
\beta\ g_{\rm I\! P}^{(3)}(\beta,Q^2_0) &=& 2.06\ \delta (1- \beta).  
\label{68} 
\end{eqnarray}

\noindent{\bf Fit~3}: Parametrizations obtained with the REN flux
factor and a initial combination of the type {\em hard--hard}:
\begin{eqnarray} 
3\beta\ q_{\rm I\! P}^{(3)}(\beta,Q^2_0) &=& 5.02\ \beta\ (1- \beta), 
\nonumber \\ 
\beta\ g_{\rm I\! P}^{(3)}(\beta,Q^2_0) &=& 0.98\ \beta\ (1- \beta).  
\label{70}
\end{eqnarray}

All these three combinations of flux factors and parton distributions
of the Pomeron were applied in the calculation of the charm contribution 
to DDIS structure function. This quantity is given by the formula
\begin{equation}
F_2^{(c)}(\beta,Q^2,x_{\rm I\! P}) = f_{\rm I\! P/p}(x_{\rm I\! P})\  
F_{\rm I\! P}^{(c)}(\beta,Q^2),
\label{72}
\end{equation}
where $ f_{\rm I\! P/p}(x_{\rm I\! P})$ stands for the integrated (over $t$) 
flux factors mentioned above and the charm structure function of the Pomeron 
is defined by Eqs.~(\ref{52})-(\ref{56}).

The results of our calculations are shown in Figs.~1-3. The upper curve
in each figure corresponds to the total diffractive structure function, 
$x_{\rm I\! P} F_2^{D}$, while the lowest one describes its charm component, 
$x_{\rm I\! P} F_2^{(c)}$. The difference $x_{\rm I\! P} 
(F_2^{D} - F_2^{(c)})$ is also shown. 

In these figures, the theoretical results are presented together with 
recent H1 and ZEUS data on $F_2^{D}$ \cite{H1new,ZEUSnew} which were 
not used in the 
fitting procedure mentioned above. The idea is not providing a precise
description for these data, but giving the reader a possibility to  
compare the net charm contribution to the precision of present-day data.

As one can see in Figs.~1-2, the charm contribution to the diffractive 
structure function obtained with the STD flux factor amounts to 30\% - 40\%,  
depending on the values of 
$\beta, x_{\rm I\! P}$, and $Q^2$. To compare, the non-diffractive structure 
function $F_2$ contains between 10\% (low $Q^2$) and 30\% (high $Q^2$) of 
charm at small $x$. From these figures, we see that the charm contribution 
to $F^D_2$ grows with the decrease of $x_{\rm I\! P}$ and is a little 
bit larger for the hard gluon distribution (Fig.~1) than it is for the 
super-hard gluons (Fig.~2). However, for both parametrizations (Fig.~1 
and Fig.~2) it is comparable with the experimental 
errors\footnote{Statistical and systematic errors have been added in 
quadrature.} of the H1 and ZEUS data 
and, consequently, can likely be measured in forthcoming HERA experiments 
on diffractive dissociation processes. 

On the other hand, for the renormalized flux factor the charm component
is very small in the full range of $\beta$ and $Q^2$ presented in Fig.~3.
The reason is that the initial gluon distribution for this case, 
Eq.~(\ref{70}), is much smaller as compared to the initial gluon 
distribution with the same form for the standard flux factor, 
Eq.~(\ref{66}).

Another way of comparing these results is shown in Fig.~4 in terms of 
$F_{\rm  I\! P}^{(c)}(\beta,Q^2)$, which is  calculated for the three 
combinations of Pomeron flux factors with the respective structure 
functions considered here.

Two main features that characterize our predictions for 
$F_{\rm  I\! P}^{(c)}(\beta,Q^2)$ are evident in this figure: 
(1) the shape of the $\beta$ distributions are quite similar (they are 
moderately hard at the initial scale) and change similarly with $Q^2$
evolution; (2) the amount of charm is different in each case 
with the proportions seen in the figure.

\subsection{Comparison with other models}

The diffractive production of the open charm in DIS has been 
studied in the framework of perturbative two--gluon exchange between the 
$c \bar c$--pair and the proton in Refs.~\cite{Genovese}. In 
Refs.~\cite{Deihl}, non--perturbative approaches were used to calculate 
cross sections and spectra for charm quark pair production. 

One common aspect of some of these models (the first two of  
Refs.~\cite{Genovese} and the first one of Ref.~\cite{Deihl}) which is in
contrast with the results of our analysis shown in Fig.~4 is that their 
predictions for 
the charm contribution practically do not change at low $\beta$ with 
$Q^2$ evolution. Another distinctive feature of these models in respect
to ours is that the $\beta$ distributions are generally peaked at some
intermediate $\beta$ value that becomes larger with increasing $Q^2$. 
This last aspect is also observed in the analysis by Levin {\it et al.} 
\cite{Genovese}, although in this case the low $\beta$ behavior does not
follow the others.

Another general observation is that the obtained steep rise of the charm 
component towards small $x_{\rm I\! P}$
is in qualitative accordance with the results of Refs.~\cite{Genovese}.

In Fig.~5, we present a quantitative comparison of our results for 
the charm contribution to $F_2^{D(3)} (x_{\rm  I\! P}, \beta,Q^2)$ 
with of those obtained by Lotter \cite{Genovese} for two $Q^2$ values and 
$x_{\rm  I\! P} = 0.001$. It is seen that, in terms of the amount of charm, 
Lotter's predictions are comparable only to our renormalized case (Fit 3), 
although in terms of shape these distributions are quite different.
Let us note, however, that Lotter's model is not adequate to describe 
diffraction in the complete $\beta$-range as was mentioned  by the 
author~\cite{Genovese}.

Now let us consider other models, reminding that our analysis was performed 
in the context of the Ingelman--Schlein model. 
Predictions for the charm contribution to the Pomeron structure function 
have been made by using the same scheme in Refs.~\cite{Gehrmann}. However, 
no estimates of the charm contribution to the diffractive structure 
function $F_2^{D(3)}$ have been presented.

In Fig.~6, we present a comparison of our predictions for 
$F_{\rm  I\! P}^{(c)}(\beta,Q^2)$ with those obtained by Haakman {\it et al.} 
\cite{Gehrmann}. We see that in their analysis the charm structure 
function is pretty soft even at low $Q^2$ where our results are predominantely 
hard. In terms of amount of charm, their results are comparable only to
our Fit 3.

\section{Concluding remarks}

We have considered in this paper the charm content of the Pomeron
and its effects on the structure function measured in diffractive deep 
inelastic scattering. 

In the present analysis, the formulas are 
derived in a way to define this contribution from the quark and gluon 
distributions inside the Pomeron obtained previously by fitting HERA 
data on diffractive deep inelastic scattering. Two parametrizations have 
been chosen for the standard Pomeron flux factor corresponding to the 
hard and superhard gluon components of the Pomeron, whereas for the 
renormalized flux factor, the hard parton parametrization has been 
analyzed.

Numerical calculations show that the results depend crucially on 
the Pomeron flux factor. In particular, the charm content of the 
Pomeron is expected to be very small for the renormalized flux factor.  
As for charm contribution corresponding to the standard flux factor, 
the estimates obtained allow us to think that it could be extracted from  
diffractive deep inelastic process with open charm production, taking
into account the planned upgrades of the HERA experiment~\cite{Eichler}.

\section*{Acknowledgements}
We would like to thank the Brazilian governmental agencies CNPq and 
FAPESP for financial support. One of us (A.V.K.) is indebted to the
Departamento de Raios C\'osmicos e Cronologia of the Instituto de 
F\'{\i}sica {\em Gleb Wataghin} (UNICAMP) for its hospitality and 
support during the course of this work. He also acknowledges helpful 
conversations with A.B.~Kaidalov.

\vfill \eject

\newpage

\centerline{\bf Figure Captions}

\noindent Fig.~1 - Theoretical estimations of diffractive DIS structure 
functions in comparison with HERA data. The curves correspond to the 
total diffractive structure function, $x_{\rm I\! P} F_2^{D}$ 
(solid curves), its charm component, $x_{\rm I\! P} F_2^{(c)}$ (dotted 
curves), and the difference $x_{\rm I\! P} (F_2^{D} - F_2^{(c)})$ 
(dashed curves). In the theoretical calculations were employed the  
flux factor and initial parton distributions of the Pomeron 
described as Fit 1 (see text). The experimental data are from 
H1~\cite{H1new} (filled circles) and ZEUS~\cite{ZEUSnew} (open circles) 
collaborations.

\vspace{1cm}

\noindent Fig.~2 -  The same as Fig.~1, but with theoretical curves 
calculated from the formulas of Fit 2 (see text).

\vspace{1cm}

\noindent Fig.~3 -  The same as Fig.~1, but with theoretical curves 
calculated from the formulas of Fit 3 (see text).

\vspace{1cm}

\noindent Fig.~4 -  Predictions for the charm structure function 
in diffractive DIS as obtained with the parametrizations of Fit 1 (a), 
Fit 2 (b), and Fit 3 (c) and their respective $Q^2$ evolution.

\vspace{1cm}

\noindent Fig.~5 -  Comparison of the charm contribution to 
$F_2^{D(3)} (x_{\rm  I\! P}, \beta,Q^2)$ obtained in the present analysis
(Fits 1, 2, and 3) with the predictions by Lotter \cite{Genovese} 
for two $Q^2$ values and for $x_{\rm  I\! P} = 0.001$.

\vspace{1cm}

\noindent Fig.~6 -   Comparison of the charm structure function obtained 
in the present analysis (Fits 1, 2, and 3) with the predictions by 
Haakman {\it et al.} \cite{Gehrmann}.

\end{document}